\documentstyle[seceq,preprint,epsf]{jpsj}

\def\runtitle{Uniaxial-Pressure induced Ferromagnetism}
\def\runauthor{Shin-Ichi {\sc Ikeda} $et$ $al.$}

\title
{
Uniaxial-Pressure induced Ferromagnetism of \\
Enhanced Paramagnetic Sr$_3$Ru$_2$O$_7$
}

\author
{ 
Shin-Ichi {\sc Ikeda}\footnote{E-mail: ikeda-shin@aist.go.jp}, Naoki {\sc Shirakawa}, Takashi {\sc Yanagisawa},\\
Yoshiyuki {\sc Yoshida}$^{1,}$\footnote{also at Nanoelectronics Research Institute, AIST, 
Tsukuba,Ibaraki 305-8568, Japan}, Shigeru {\sc Koikegami}$^{1}$, Soh {\sc Koike}$^{1}$,\\
Masashi {\sc Kosaka}$^{2}$, and Yoshiya {\sc Uwatoko}$^{3}$
}

\inst
{
Nanoelectronics Research Institute, AIST, Tsukuba, Ibaraki 305-8568, Japan\\
$^1$JSPS, Chiyoda, Tokyo 102-8471, Japan\\
$^2$Department of Physics, Saitama University, Saitama 338-8570 , Japan\\
$^3$ISSP, University of Tokyo, Kashiwa, Chiba 277-8581, Japan
}

\recdate
{
\today
}

\abst
{
We report a uniaxial pressure-dependence of magnetism in layered perovskite 
strontium ruthenate Sr$_3$Ru$_2$O$_7$. By applying a relatively small uniaxial pressure,
greater than 0.1 GPa normal to the RuO$_2$ layer, ferromagnetic ordering manifests
below 80 K from the enhanced-paramagnet. Magnetization at 1 kOe and 2 K becomes 
100 times larger than that under ambient condition. Uniaxial pressure dependence of Curie
temperature $T{\rm_C}$ suggests the first order magnetic transition. Origin of this 
uniaxial-pressure induced ferromagnetism is discussed in terms of the rotation of RuO$_6$ 
octahedra within the RuO$_2$ plane.
}

\kword
{
Sr$_3$Ru$_2$O$_7$, Sr$_2$RuO$_4$, ferromagnetism, uniaxial pressure, nearly ferromagnetic metal, magnetization, structural distortion 
}

\begin{document}
\sloppy
\maketitle

\section{Introduction}

Tuning the magnetism in solid state compounds via tiny perturbation is one 
of the central issues for strongly-correlated electron systems such as 
heavy-fermion (HF) intermetallic compounds.  Since those materials tend to reveal 
the high susceptibility of the electronic properties to relatively small 
external pressures ($\leq$ GPa). For instance, antiferromagnetic HF 
compounds (CeCu$_2$Ge$_2$ \cite{rf:1}, CeRh$_2$Si$_2$
\cite{rf:2}, CePd$_2$Si$_2$
\cite{rf:3})
and a ferromagnetic HF compound (UGe$_2$ \cite{rf:4}) reveal superconductivity under several GPa 
hydrostatic pressures, which reduce the magnetic ordering 
temperature towards absolute zero. Around the diminishing region, the superconductivity 
appears. Several GPa pressures, which is necessary to induce the superconductivity 
in above HF compounds, are relatively small for solid considering that 
much more pressures are probably required to suppress ferromagnetic ordering temperatures
above 600 K completely in conventional ferromagnets like Fe. In fact, under
hydrostatic pressure,
ferromagnetic Fe undergoes a structural phase transition at around 10 GPa.
Below 10 GPa, ferromagnetic phase remains without showing large
reduction of the ferromagnetic transition temperature \cite{rf:5,rf:6,rf:7}.
Concerning pressure induced magnetism from paramagnetism, Umeo $et$ $al$. reported 
that uniaxial pressure induced antiferromagnetism in CeNiSn 
(paramagnetic at ambient) above 0.1 GPa at 6 K \cite{rf:8}.

In the present study, we report a uniaxial pressure induced paramagnetic to 
ferromagnetic transition in two-dimensional (2D) ruthenium oxides (ruthenates) 
Sr$_3$Ru$_2$O$_7$ characterized by a paramagnetic state at ambient
pressure \cite{rf:9}. Hydrostatic pressure is not vital to this transition
as described later. 
Ruddelsden-Popper type ruthenates A$_{n+1}$Ru$_n$O$_{3n+1}$ (A: divalent alkali metal such 
as Sr and Ca, $n$ = 1, 2, 3 and infinity) exhibit rich and comprehensive electronic 
ground states. The spin-triplet superconductor \cite{rf:10,rf:11,rf:12}, ferromagnetic
metal \cite{rf:13}, 
antiferromagnetic Mott insulator \cite{rf:14} and antiferromagnetic
metal \cite{rf:15} have been 
investigated extensively. Sr$_3$Ru$_2$O$_7$, which has a 2D crystal structure 
(Fig.1), is another such compound \cite{rf:16,rf:17,rf:18,rf:19,rf:20}. 

The first report on single crystals grown 
via a floating-zone method indicated that Sr$_3$Ru$_2$O$_7$ is a paramagnet that exhibits 
enormously strong ferromagnetic instability, which is most consistent with the 
results of polycrystals \cite{rf:16,rf:17,rf:19}. Temperature dependence of magnetic susceptibility 
$\chi$($T$) is nearly isotropic and obeys Curie-Weiss law with antiferromagnetic Weiss temperature above 200 K.
At around 17 K, $\chi$($T$) shows a broad maximum which is often observed in nearly ferromagnetic 
(enhanced paramagnetic) metal like Pd or TiBe$_2$. Below 7 K, temperature independent $\chi$($T$) and
$T$-square behavior of in-plane and out of plane resistivities are recognized, indicating the Fermi liquid
state with large Wilson ratio $R{\rm_W}$ $\geq$ 10. Field dependence of magnetization $M(H)$ below 17 K reveals
metamagnetic transitions around 50 kOe for $H||ab$ and 80 kOe for $H||c$. High purity single crystalline
Sr$_3$Ru$_2$O$_7$ with in-plane residual resistivity of 0.8 $\mu \Omega$ cm does not exhibit superconductivity
down to 200 mK. Very recently, Grigera
$et$ $al$. have reported the possibility 
of quantum critical phenomena at metamagnetic transition in
Sr$_3$Ru$_2$O$_7$ \cite{rf:21}. The quantum 
criticality around the expected critical point corresponding to the first-order 
metamagnetic transition is probably related to the strong ferromagnetic
instability \cite{rf:22,rf:23,rf:24}. 
Several materials exhibiting strong ferromagnetic instability are recognized as 
being nearly ferromagnetic paramagnets. However, Sr$_3$Ru$_2$O$_7$ is the only compound of 
the nearly ferromagnetic paramagnet with 2D crystal structure.

\begin{figure}
\leavevmode
\begin{center}
\epsfxsize=50mm
\epsfbox{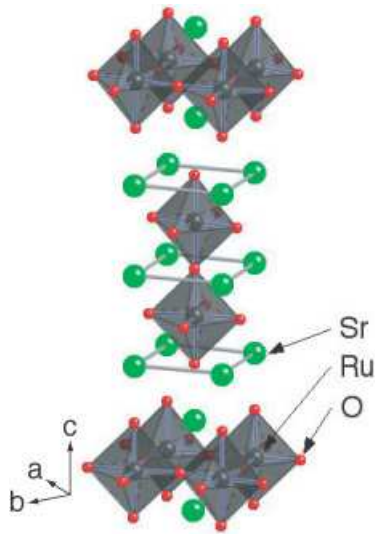}
\end{center}
\caption{Crystal structure of bilayered strontium 
ruthenate Sr$_3$Ru$_2$O$_7$. Strontium and ruthenium ions are positioned at 
tetragonal symmetry sites characterized by the space group of I4/mmm, 
which is the same as that of superconducting single-layered
Sr$_2$RuO$_4$. 
SrO rock-salt layers and RuO$_2$ layers are alternately stacked forming 
a typical layered perovskite structure with RuO$_6$ octahedra. Each
RuO$_6$
octahedron shares its oxygens at the corner. Considering oxygen ions,
the symmetry is lowered to orthorhombic space group of $Bbcb$ due to 
the RuO$_6$ rotation by about 7 degrees. The c-axis is normal to the 
two-dimensional RuO$_2$ planes.}
\end{figure}

In order to amplify the ferromagnetic instability in Sr$_3$Ru$_2$O$_7$, we performed magnetization
measurements under hydrostatic pressure and observed the sign of ferromagnetic ordering 
at approximately 1 GPa \cite{rf:9}. However, improved technique proved that hydrostatic pressures 
up to 1.4 GPa do not induce magnetic ordering \cite{rf:25}. Therefore, considering the possibility 
of a pressure component that is uniaxial with respect to the ferromagnetism, we performed 
uniaxial pressure measurements on single-crystalline Sr$_3$Ru$_2$O$_7$ for the first time. Uniaxial 
pressure was found to induce paramagnetic-to-ferromagnetic transition at pressures over 
0.1 GPa, which, for inorganic materials, is a rather low critical pressure.

\section{Experimental method}
Single crystalline Sr$_3$Ru$_2$O$_7$ was grown using a typical floating-zone technique
\cite{rf:26}. Magnetization measurements under uniaxial pressure were performed by SQUID 
 (superconducting quantum interference device) magnetometer. The uniaxial pressure 
 ($p$ $\leq$ 0.5 GPa) and magnetic field ($H$ $\leq$ 55 kOe) were applied along the normal axis (c-axis)
 to the 2D RuO$_2$ plane in Sr$_3$Ru$_2$O$_7$. The pressure is estimated based on the shift of the 
 superconducting transition temperature $T_{\rm c}$ of Sn
 \cite{rf:27}.

\section{Results and Discussion}
Figure 2 illustrates the field-dependent magnetization $M$($H$) for several uniaxial 
 pressures up to approximately 0.5 GPa at 2 K. Linear field dependence is observed 
 for ambient atmospheric conditions (i.e. uniaxial pressure $p$ = 0 GPa), which is consistent
 with paramagnetic behavior. 

 \begin{figure}
\leavevmode
\begin{center}
\epsfxsize=80mm
\epsfbox{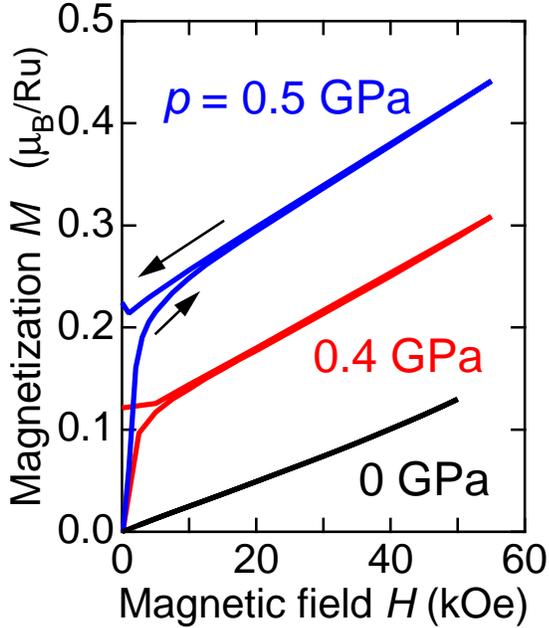}
\end{center}
 \caption{Magnetic field dependence of magnetization 
 $M$($H$) of single-crystalline Sr$_3$Ru$_2$O$_7$ at the temperature of 2 K. 
 Magnetic field and uniaxial pressure are applied along the c-axis. 
 Uniaxial pressure above 0.1 GPa causes ferromagnetic phase from paramagnet. 
 The observed linear field dependent $M$($H$) reflects paramagnetic phase 
 at $p$ = 0 GPa. Increasing pressure leads to a larger remanent magnetic 
 moment at zero field. The hysteresis curves, which do not saturate at 
 the field of 55 kOe, are typical for metallic magnetism with a 
 ferromagnetic component below the saturation field. }
\label{fig:2}
 \end{figure}

 At $p$ $\geq$ 0.1 GPa, ferromagnetic characteristics appear unexpectedly in the magnetization curve $M$($H$),
 indicating pressure-induced ferromagnetic transition from the paramagnet phase. With
 pressures increasing beyond this critical value, the magnetization is enhanced. At 
 $p$ = 0.5 GPa and $H$=55 kOe, magnetization reaches almost 0.5 $\mu_{\rm B}$/Ru. In addition, no 
 saturation behavior is observed at higher fields. This type of non-saturated feature 
 is typical for metallic ferromagnets such as SrRuO$_3$ under the magnetic field smaller than the saturation field \cite{rf:28}. 

 Assuming fully localized 4d electron in Ru ion sites (Ru$^{4+}$) and a stronger crystalline 
electrical field than spin-orbit coupling, four 4d electrons are distributed in three 
$t_{\rm 2g}$ orbitals ($d_{xy}, d_{yz}, d_{zx}$), resulting in spin quantum number 
of $S$=1, corresponding 
to 2$\mu_{\rm B}$/Ru. If the uniaxial pressure-induced ferromagnetic transition in the present study 
is canted antiferromagnetic, the canting angle of Ru spins should be greater than 14 degrees 
in the case of completely localized electron phases. Such a localized system implies the 
insulating conductivity phase known as a Mott insulator. Electrical resistivity measurements
under uniaxial pressure along c-axis in Sr$_3$Ru$_2$O$_7$ up to a pressure of 0.1 GPa imply that 
there is no metal-insulator transition \cite{rf:29}. This indicates that uniaxial pressure-induced 
ferromagnetic transition does not concur with charge transport transition.

In the uniaxial pressure-induced magnetic phase,  
$S$ can not easily 
be assumed to be approximately 1. In itinerant electron
systems, $S$ is much smaller than that 
in the localized case. Hence, the canted antiferromagnetic phase in Sr$_3$Ru$_2$O$_7$ is possible 
only when the canting angle of Ru-spins is much greater than 14 degrees. 
It is appropriate to assume a finite spin-orbit ($L-S$) coupling in these ruthenates due to 
the large mass of Ru ions. In such a case, structural modifications are related to the spin 
canting angle. Huang $et$ $al$. and Shaked $et$ $al$. reported that the RuO$_6$ octahedron in Sr$_3$Ru$_2$O$_7$ rotates 
about the c-axis by 7 degrees \cite{rf:30,rf:31}. Other structural strains, such as tilting
from the 2D plane, were not 
detected. If the rotation angle matches the spin canting angle due to the finite 
$L-S$ coupling, 
the spin canting in the uniaxial-pressure-induced magnetic phase in Sr$_3$Ru$_2$O$_7$ exists within 
the 2D RuO$_2$ plane and its angle is approximately 7 degrees. However, the canting angle of 
7 degrees within the RuO$_2$ plane can not explain the large moment of approximately 0.5 $\mu_{\rm B}$/Ru 
in the metallic phase as discussed above. In addition, spin canting within the 2D plane 
cannot induce a ferromagnetic moment along the c-axis in this crystal structure. Therefore, 
the observed pressure-induced ferromagnetic phase is thought to have originated from a simple
ferromagnetic alignment of Ru spins.

The temperature dependence of magnetization $M$($T$) for various uniaxial pressures is shown 
in Fig.3. Data for both field-cooling sequences and zero field-cooling sequences is presented.
Above 0.1 GPa, considerable abrupt magnetization develops at around 80 K. 
Increasing the pressure, the magnetic moment at 2 K becomes larger without changing 
ferromagnetic ordering temperatures $T_{\rm c}$. The magnetization at approximately 0.5 GPa in 
Sr$_3$Ru$_2$O$_7$ 
is 100 times greater than that at ambient pressure. Releasing the pressure to zero, 
$M$($H$) and 
$M$($T$) again become paramagnetic, indicating good reproducibility.
This enormous uniaxial
pressure effect suggests that Sr$_3$Ru$_2$O$_7$ exhibits $piezo-ferromagnetism$, which means uniaxial-
pressure induced ferromagnetism, perhaps due to the change in crystal structure explained below. 
Concerning the uniaxial pressure dependence of $T_{\rm c}$, it seems that the first order transition 
driven by the pressure is accompanied by the appearance of the ferromagnetism. 

\begin{figure}
\leavevmode
\begin{center}
\epsfxsize=80mm
\epsfbox{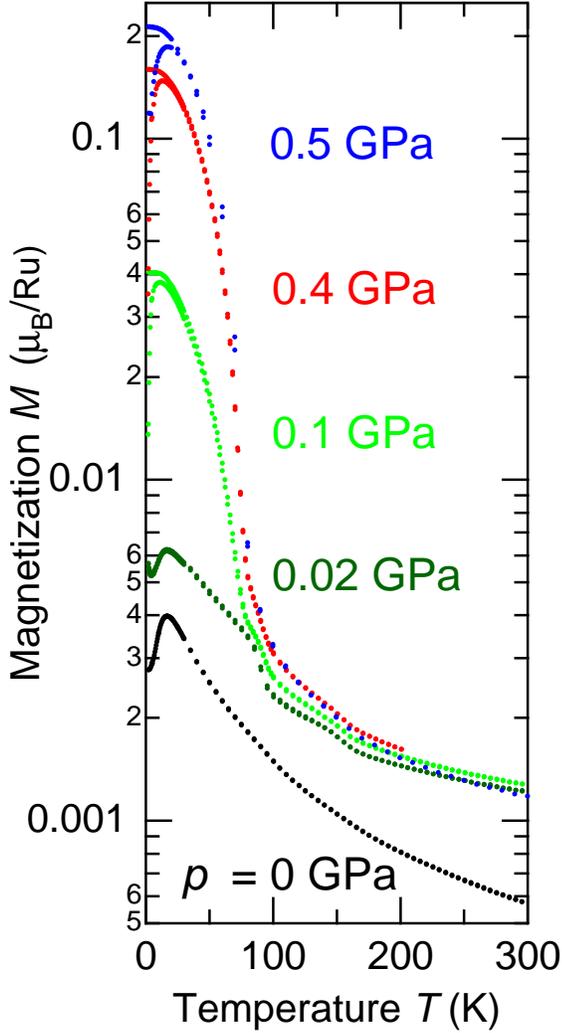}
\end{center}
\caption{Temperature dependence of magnetization $M$($T$) 
of single-crystalline Sr$_3$Ru$_2$O$_7$ for each uniaxial pressure at the field 
of 1 kOe. Magnetic field and uniaxial pressure are applied along the 
c-axis. Anomalous evolution of ferromagnetism with increasing uniaxial 
pressure is shown using a semilogarithmic scale. The ferromagnetic 
transition temperature $T_{\rm c}$ remains constant with varying the uniaxial pressure.}
\label{fig:3}
\end{figure}

In order to understand this outstanding instability of magnetism to small uniaxial pressure, 
we focus on the RuO$_6$ rotation angle in Sr$_3$Ru$_2$O$_7$. The significance of the RuO$_6$ rotation is evident 
since the observed variation in the rotation
angle with decreasing temperature in Sr$_3$Ru$_2$O$_7$ is quite large
($\approx$ 15 $\%$) in comparison with 
the case of other lattice parameters ($\leq$ 1 $\%$)\cite{rf:31}. 
On the contrary, hydrostatic pressure, which
 is not essential for the above induced-ferromagnetism, does not modify the
 RuO$_6$ rotation 
 angle \cite{rf:31}. In 2D R-P ruthenates, the rotation angle is vital in 
 determining the electronic states \cite{rf:32,rf:33}. For instance, a similar situation is realized
 in the surface state of 
Sr$_2$RuO$_4$, whereby the RuO$_6$ octahedron also rotates by 7 degrees, 
as is the case in bulk state Sr$_3$Ru$_2$O$_7$. Surface ferromagnetism caused by the rotation was 
theoretically derived \cite{rf:34}. Furthermore, considering other precious metal oxides, 
IrO$_6$ octahedra in layered perovskites Sr$_2$IrO$_4$ and Sr$_3$Ir$_2$O$_7$ also rotate along the direction normal to the 2D plane 
\cite{rf:35,rf:36,rf:37,rf:38}. 
This type of structural distortion generally dominates the electronic state in layered perovskite maerials,
by tuning electron transfer within the 2D plane. For instance, both iridates above are ferromagnetic semiconductors
reflecting the stronger effective correlation. On the other hand, Sr$_2$RuO$_4$ and Sr$_2$MoO$_4$, which have no such structural distortions, 
are Pauli paramagnetic conductors \cite{rf:39,rf:40,rf:41}.

Therefore, we regard the rotation angle as an 
 indicator of the uniaxial-pressure-induced ferromagnetism. 
Our calculations based on the 2D three-band Hubbard model with the
RuO$_6$ octahedron rotation
suggest that the quantum phase transition toward the ferromagnetic phase from quantum 
disordered state ($p$ = 0 GPa) due to Ru-4d orbital degeneracy can occur by small changes in the
RuO$_6$ 
rotation angle in Sr$_3$Ru$_2$O$_7$ \cite{rf:42}.  
These calculation results and the existence of the same
structural distortions in layered perovskite precious metal oxides may assure that the rotation angle of RuO$_6$ octahedra
is the clue to understand the pressure-induced ferromagnetism in Sr$_3$Ru$_2$O$_7$.
It is strongly required to perform the neutron diffraction measurements under the uniaxial
pressure to determine the relation between structural distortions and
magnetism  on single crystalline Sr$_3$Ru$_2$O$_7$.

\section{Conclusion}
In conclusion, we have reported the observation of remarkable uniaxial-pressure-induced 
ferromagnetic transition from the enhanced paramagnetic state in Sr$_3$Ru$_2$O$_7$.
The uniaxial pressure dependence of $T_{\rm c}$ indicates that the appearance of ferromagnetism
is corresponding to the first order phase transition driven by the pressure.
 In addition, Sr$_3$Ru$_2$O$_7$ is the candidate for the practical $piezo-ferromagnet$ due to 
its rather small critical pressure. The origin of appearance of the ferromagnetism is 
probably associated with the change in RuO$_6$ rotation angle, suggesting a possible criterion
which may be used to find further examples of practical $piezo-ferromagnet$ candidates 
with higher critical temperatures.

\section*{Acknowledgements}
The authors would like to thank H. Eisaki, A.P. Mackenzie, I. Nagai, Y.Maeno, M.K. Crawford, 
    M. Takigawa, C.H.Lee and J. Itoh for 
    their helpful advice and kind assistance. This study was 
    supported in part by a Grant-in-Aid for the Scientific Research 
    on Priority Area "Novel Quantum Phenomena in Transition Metal 
    Oxides" from the Ministry of Education, Culture, Sports, Science 
    and Technology.

\end{document}